\documentclass[draftclsnofoot,onecolumn]{IEEEtran}
\ifCLASSINFOpdf
  \usepackage[pdftex]{graphicx}
  \graphicspath{{images/}}
  \DeclareGraphicsExtensions{.pdf,.jpeg,.png}
\else
\fi
\usepackage{color}

\usepackage[cmex10]{amsmath}

\usepackage[font={small}]{caption}
\usepackage{url}

\newcommand*{\TitleFont}{%
      \usefont{\encodingdefault}{\rmdefault}{b}{n}%
      \fontfamily{ptm}
      \fontsize{14}{16}%
      \selectfont}
\begin{document}
%
\title{\TitleFont{Facebook Applications’ Installation and Removal: A Temporal Analysis}}


\author{\IEEEauthorblockN{Dima Kagan,
Michael Fire,
Aviad Elyashar,
and Yuval Elovici\\
}
\IEEEauthorblockA{ Telekom Innovation Laboratories and Information Systems Engineering Department, \\
Ben-Gurion University of the Negev,  Beer-Sheva, Israel\\
 Email: \{kagandi,mickyfi,aviade, elovici\}@bgu.ac.il}
}


%


\maketitle

\begin{abstract}
Facebook applications are one of the reasons for Facebook attractiveness.
Unfortunately, numerous users are not aware of the fact that many malicious Facebook applications exist.
To educate users, to raise users’ awareness and to improve Facebook users’ security and privacy, we developed a Firefox add-on that alerts users to the number of installed applications on their Facebook profiles.
In this study, we present the temporal analysis of the Facebook applications’ installation and removal dataset collected by our add-on.
This dataset consists of information from 2,945 users, collected during a period of over a year. 
We used linear regression to analyze our dataset and discovered the linear connection between the average percentage change of newly installed Facebook applications and the number of days passed since the user initially installed our add-on. 
Additionally, we found out that users who used our Firefox add-on become more aware of their security and privacy installing on average fewer new applications. 
Finally, we discovered that on average 86.4\% of Facebook users install an additional application every 4.2 days.

\end{abstract}



%
\IEEEpeerreviewmaketitle

\section{Introduction}
In the last decade, online social networks have gained enormous popularity.
Over a billion users worldwide are using these networks \cite{itu2012trends} to share information, communicate with friends, play games, etc. \cite{fac}. 
Recently, J.B. Duggan \cite{duggan2013demographics} discovered  that 67\% of adults in the United States use online social networks. These results demonstrate a 2\% increase in comparison to a previous report  \cite{2012demographics}.  
The exponential growth of social networks created a reality where there are social networks for almost every usage. These vary from social networks for connecting with business colleagues like LinkedIn \cite{linkedin} and Xing \cite{xing} to social networks for animal lovers, such as Dogster \cite{dogster}, Catster \cite{catster} and YummyPets \cite{yummypets}.
The biggest online social network is Facebook, \cite{facebook} which has more than 1.11 billion monthly active users as of March 2013 \cite{fac}. 
The median Facebook user is 22 years \cite{met}, and has 138 friends on average \cite{fac}.
Facebook users have made 140.3 billion friend connections and used over 1.13 trillion likes \cite{met}. 
Moreover, every 60 seconds Facebook users post 510,000 comments, update 293,000 statuses, and upload 136,000 photos \cite{zephoria}. 
One of the main reasons for Facebook popularity is its third-party applications \cite{rahman2012frappe}. 
The Facebook application platform popularity is growing rapidly; for example, more than 20 million Facebook applications are installed every day \cite{car}.  
There are many different kinds of Facebook applications,  for example there are utility, productivity, and even educational applications \cite{fbappcenter}.
According to Nazir et al. \cite{nazir2012beyond}, the most popular applications are games; approximately 230 million people play games on Facebook every month \cite{gre}. The most popular games such as ``Candy Crush Saga'' have more than 2.7 million daily active users \cite{ing}. \\
\indent In recent years, hackers and spammers have found Facebook applications to be an efficient platform for spreading malware and spam. Moreover, recent research has found that at least 13\% of applications on Facebook are malicious \cite{rahman2012frappe}. Recently, Rahman et al. \cite{rahman2012frappe} described in their study that spammers use Facebook applications to lure their victims into clicking on specific malicious links.
Additionally, hackers can take advantage of Facebook application platform properties to: (a) find their next potential targets across a large base of potential users \cite{rahman2012frappe}, (b) use the trust between friends to infect more users \cite{patsakis2009social}, (c) exploit  the application developers API \cite{apifac} to collect information, such as personal information, photos, tags, posts, chat threads, etc., and (d) use the user's credentials to publish posts or to spread spam, advertising and phishing under the name of a legitimate user \cite{gao2010detecting}.\\
\indent  To deal with this problem, we developed an add-on that is part of the Social Privacy Protector (SPP) \cite{fire2012social,fire2013friend}. The purpose of this add-on is to educate and to increase user awareness about the threat that lurks in social applications. 
Our add-on notifies users of their current number of applications. By notifying the users, we encourage removal of third-party applications that are not in use by the user.  \\
\indent In this study, we utilize our dataset which was collected by our dedicated Firefox add-on. It contains information about 44,541 different occasions collected from   2,945 users between May 2012 and June 2013. 
Each entry in our dataset consists of the user id, application number and date. 
\\ 
\indent In our previous study \cite{fire2013friend}, we presented preliminary statistics on this dataset. 
We analyzed this dataset and discovered that users who used the SPP add-on for application removal, removed more than 50\% of all their installed applications one day after its installation. These results indicate that in many cases the installed applications are unwanted or unneeded applications.
In this study, we perform a temporal analysis of the installed application data that was collected for a longer period of time red than in our previous study.
We discovered that within the first week after the add-on's initial use, the user's number of applications decreased by 12.1\% (see Table \ref{table:appchange}) on average.
Moreover, the application removal rate continued to grow up to 27.7\% (see Table \ref{table:appchange}) by an average of 63 days after the initial use.
According to the results presented in this study, we can conclude that using our add-on made many users become more aware of the existence of unnecessary applications on their Facebook profiles.  \\
\indent The remainder of this paper is organized as follows. 
In Section II, we give a brief overview on different related solutions that help users in social networks protect themselves from malicious applications. 
In Section III, we describe our methodology for analyzing the information and how we used linear regression to predict the application number. In Section IV, we present our study's initial results.
Finally, in Section V, we present our conclusions regarding the change in awareness that resulted from our add-on notifications and offer future research directions.

\section{Related Work}
Recent reports \cite{threat2013sophos,threat2013websense} have indicated  that due to the growth of social network popularity, there  also has been a massive rise in malicious activity and security threats  to online social network users.
In recent years, social network users, social network operators, security companies, and academic researchers have proposed solutions to increase the security and privacy of social networks users. In the remainder of this section, we describe notable solutions in the area.

\subsection{Detecting Malicious Applications }
Detection is the most standard way to deal with security and privacy problems. 
There are many works in the area and many different ways to detect malware. For example, Rahman et al. \cite{rahman2012efficient} presented MyPageKeeper, a Facebook application that protects Facebook users from socware.
MyPageKeeper is based on a Support Vector Machine (SVM) classifier that uses a main feature specific keyword occurrence in a post made by an application. MyPageKeeper was able to identify socware posts and alert the user with 97\% accuracy, but was unable to detect malicious applications. 
Websense Defensio \cite{abu2011malicious} is a Facebook application from Websense that monitors posts in a user’s profile and determines whether they are legitimate, spam, or malicious. Defensio also uses SVM to detect malicious posts and in addition they could delete them. Abu-Nimeh et al. \cite{abu2011malicious} used Defensio as a platform to study malicious links. They found that about 9\% of the studied posts were spam or malicious.
In 2012,  Rahman, et al. \cite{rahman2012frappe} improved his previously mentioned work. Rahman, et al. developed the FRAppE: A tool that can identify malicious applications by using the application information as features. Some examples include the number of permissions required, the domain reputation of redirect URI, and others.
FRAppE can detect malicious applications with 99.5\% accuracy and a low false negative rate 4.1\%.

\subsection{Increasing User Awareness of the Threat}
Another approach is informing the user about possible threats
that can jeopardize his or her security and privacy. By using this approach, it is possible to stimulate users to react and to protect themselves. This tactic is about teaching users what kind of threats exist and how they should protect themselves.\\
\indent In 2011 Wang et al. \cite{wang2011third} suggested to change the way that permission are asked of the user; for example: by alerting the user that there is a conflict between his or her privacy settings and the permissions that the application requires.
In 2012, Xu et al. \cite{xuprivacy} chose to redesign the Facebook application authentication dialogue to increase user awareness of the permissions that are required by applications. 
Recently, Fire et al. \cite{fire2012social} presented the SPP software, a Firefox add-on and a Facebook application.
This software contains three protection layers, which improve user privacy by implementing different methods.  
(1) \textit{The friends layer} - suggests friends who might pose a threat and then restricts these friends’ exposure to the user’s personal information. (2) \textit{The privacy settings layer} – is based upon different types of social network usage profiles. (3) \textit{The application layer} - alerts users to the number of installed applications on their Facebook profiles. 
\\
\indent In our previous study \cite{fire2013friend}, we  presented initial results, which were based on a dataset from 1,676 users collected  between the 27th of June, 2012, and the 10th of November, 2012.
In this study, we perform a temporal analysis on our complete dataset, which contains information from 2,945 users between May 2012 and June 2013.
Unlike our previous study, this is the first study where we focused on analyzing user application installation and removal from different aspects for long periods of time.

\section{Methods and Experiments}
\begin{figure}[h]
\vspace{-15pt}
\begin{center}
\includegraphics[scale=1]{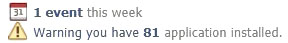}
\end{center}
\caption{Warning about installed applications.}
\label{figure:add_on}

\vspace{-16pt}
\end{figure}

Our study’s main goal is to study how awareness affects installation and removal of social applications.
To perform our research, we collected data from the Facebook online social network by using the SPP Firefox add-on (see Figure \ref{figure:add_on}), which collected the following information:
(1)\textit{ Hashed User Id},
(2) \textit{Installed Application Number} - the number of installed Facebook applications on the user’s Facebook account, and
(3) \textit{Date} - the date when the information was collected.	
To avoid duplicate entries, we collected only one entry for each user per day.
Our final collected dataset included 2,947 Facebook users with unique hashed user ids across 351 different days between the months of May 2012 and June 2013. 
On average, we obtained information of 63 days and 15.1 entries per user.
In addition, we discovered that Facebook users have a mean of 41.86 applications.\\

\indent To perform temporal analysis of our dataset we carried out the following steps.
First of all, we divided all the users into two groups as follows: (1) \textit{Regular Users} - users whose number of installed Facebook applications did not decrease between the initial and last use of the add-on (2) \textit{Add-on Users} - users who decreased the number of installed Facebook application between their initial and last use of the add-on. 
Next, for each user, we calculated $AppChangeRatio(u,d)$: the ratio between the change in the number of applications user, $u$, had after, $d$, days and the number of applications that, $u$,  had when the add-on  was initially installed.
$AppChangeRatio(u,d)$ was calculated only for the first 63 days after the add-on was initially installed, where 63 days is the average add-on use period.
The formal definition of the $AppChangeRatio$ function is as follows:
\begin{equation}
\
AppChangeRatio(u,d):= ACR(u,d):=
\frac{AppNum(u,0) \text{-} AppNum(u,d)}{AppNum(u,0)}
\end{equation}
Where  $AppNum(u,d)$ is a function that returns the number of applications user, $u$, had at a day, $d$, after the add-on installation.
Then, for each day, $d$, we calculated the average $ACR(u,d)$ for each group as follows: 
\begin{equation}
\
AvgAppChange(u,d):= AvgAC(u,d):=
\frac{\sum_{ u \in Users}ACR(u,d)}{|Users|}
\end{equation} 
We performed  \textit{t-test} to reject the null hypothesis, which states that the mean number of applications of the \textit{Regular Users} and the \textit{Addon Users} is identical  and the observed differences are merely random.
To find correlation between the application number and time, we conducted a linear regression experiment for each of the groups. The independent variable is the number of days passed since the initial install. The dependent variable is the percentage change in the applications number since the initial install. \\
\indent Lastly, we tested if there are specific days of the week when users install or remove more applications than on other days. 
To test this we calculated the average change in application number for every day of the week, $w$, for all the users, $u$. The change in application number is the calculation of difference in the number of applications between sequel days, we defined \textit{AppNumDelta} as: 
\begin{equation}
\
AppNumDelta(u,d):=ANDelta(u,d):=
AppNum(u,d)-AppNum(u,d+1)
\end{equation}
The results were divided into two cases: \\
\begin{equation}
\
I(w)= \frac{\sum_{\{wday(d) = w \mid ANDelta(u,d)>0\}} ANDelta(u,d)}{|\{u \in Users| wday(d) =  w \wedge ANDelta(u,d)>0\} |}
\end{equation}
\begin{equation}
\
R(w)= \frac{\sum_{\{wday(d) = w \mid ANDelta(u,d) \leq 0\}} ANDelta(u,d)}{|\{u \in Users| wday(d) =  w \wedge ANDelta(u,d)\leq0\} |}
\end{equation}
where $wday(d)$ returns the day of the week for $d$. $I(w)$ is the average number of applications installed on that day of the week, $w$, and $R(w)$ is the average number of applications removed on that day of the week, $w$.

\section{Results}
At the end of our analysis, our \textit{Regular Users} group  consisted of  2,545 users and the \textit{Add-on Users}  group consisted of 400 users, out of which 52 (13\%) also used our Friends Analyzer application,  which is responsible for identifying a user's friends, who may pose a threat to the user’s privacy. 
Therefore, we can classify these 52 users as privacy concerned users. Further, this suggests that less than 2\% of all Facebook users are highly aware of their privacy.\\
\begin{table}[h]\small
\vspace{-5pt}
\caption{The Percentage Change in the Applications Number} \label{table:appchange}  
\begin{center}	
    \begin{tabular}{| l | l | l | l |}
    \hline
    Group & After 1 Day & After 7 Days & After 63 Days \\ \hline
    Regular Users & 2.02\% & 11.9\% & 40.7\% \\ \hline
    Add-on Users & -5.4\% & -12.2\% & -27.7\%
     \\
    \hline
    \end{tabular}
\end{center}
\vspace{-10pt}
\end{table} 

Afterward, we performed a Welch Two Sample t-test to prove the assumption of difference between the two groups. 
The t-test confirms that the \textit{Add-on Users}  $(\mu=0.236,\; \sigma_{dev}= 0.12)$ and the \textit{Regular Users} $(\mu=-0.19,\; \sigma_{dev}=0.05)$ have a significant difference  in their mean values  $  t = 25.936,\; p-value < 2.2e-16$.
Next, we used the linear regression method to find the correlation between the  change ratio in number of applications and period of time in days. \\

\begin{figure}[h]
\vspace{-10pt}
\begin{center}
\includegraphics[scale=0.8]{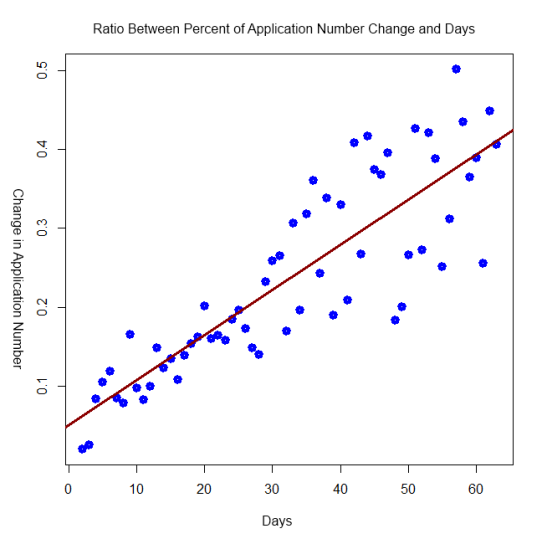}
\caption{The ratio between the percentage change in the applications number and days passed for the \textit{Regular Users}. The solid line equation - $ApplicationChangePercent=0.006×Days+ 0.05$}
\label{figure:plot_group1}
\end{center}
\vspace{-10pt}
\end{figure}

\begin{figure}[h]
\vspace{-10pt}
\begin{center}
\includegraphics[scale=0.8]{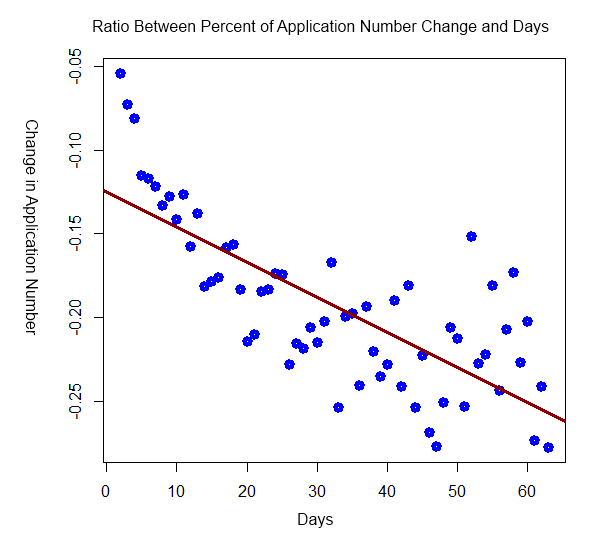}
\caption{The ratio between the percentage change in the number of applications and days passed of the \textit{Add-on Users}. The solid line equation - $ApplicationChangePercent= -0.002×Days-0.125$ }
\label{figure:plot_group2}
\end{center}
\vspace{-10pt}
\end{figure}

As a result of the difference we have proven between the two groups, we divided the linear regression into two cases: (1) \textit{Regular Users} - users whose number of applications increased or did not change (see Figure ~\ref{figure:plot_group1}), (2) \textit{Add-on Users} - Users whose  number of applications decreased (see Figure ~\ref{figure:plot_group2}).

Using linear regression, we received the following regression equations (see Figure ~\ref{figure:plot_group1} and ~\ref{figure:plot_group2}):
\begin{equation}
\
Regular Users:
ApplicationChangePercent=
0.006×Days+ 0.05
\end{equation}
where $R^2= 0.736 ,\; and \; p-value = 2.2e-16 .$
\begin{equation}
\
Addon Users:
ApplicationChangePercent=
-0.002×Days-0.125  
\end{equation}
where $R^2= 0.57 , \; and \; p-value = 1.351e-12$ \\
According to the linear regression results, the average Facebook user application number increases linearly over time, and the user installs about 7.15 applications each month.

\begin{figure}[h]
\vspace{-10pt}
    	\centering
		\includegraphics[scale=0.50]{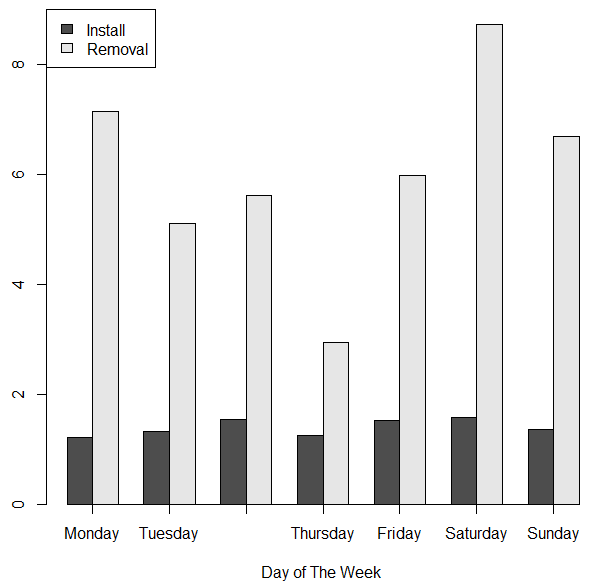}
		\caption{Average application install and removal per day of the week.}
	\label{figure:bar_day}
	\vspace{-7pt}
\end{figure}	

\indent In addition, we tested if there is a specific day of the week when users install or remove more applications than normal. 
We discovered that for both cases there is more activity on Saturdays and a significantly lower removal rate on Thursdays (see Figure ~\ref{figure:bar_day}).

\section{Conclusion}

In this study, we presented our initial methods and results in studying online social network applications with an aim of improving users safety and awareness. According to our results, it is possible to predict the number of applications a casual user has with high accuracy. 
Moreover, we discovered that users install many applications every month, and the average user installs approximately 7.15 applications each month. Currently, there are several malicious applications available. Rahman et al. states \cite{rahman2012frappe} that at least 13\% of Facebook applications are malicious. By joining these two statistics together, we shockingly conclude that on average a Facebook user installs more than ten malicious applications each year.
Our results show a possible solution to this problem. After the installation of our add-on, users become more aware of the number of installed applications they had. Users removed on average 12.1\% of their applications after the first week, and they continued to remove even more applications afterward. Furthermore, by using the equations we discovered, it is possible to classify users who are not aware of the number of applications they have.\\
\indent In addition, we discovered that users install and remove more applications on Saturdays. We assume this is due to many users being at home instead of working on this day of the week. It is possible to use these results to notify users more often, in ways that are more noticeable and on specific dates regarding the danger of installing applications. For example, we can configure our add-on to give more alerts and add special alerts on Saturday.\\
\indent The study presented in this paper is a work in progress with many available future directions.
By gathering additional information about what kind of applications users tend to restrict, we can develop an algorithm for application removal recommendations.
Moreover, when the same applications are restricted by many users, we can conclude with high likelihood that these applications are fake applications and recommend to Facebook and our users to remove these applications from the social network and their accounts. 
Another possible future direction is  discovering the point in time when the \textit{Add-on Users’}  application  numbers start increasing again, and at that point, to give the user a special warning regarding his or her number of applications.



%
\bibliographystyle{IEEEtran}
\bibliography{spp}

\end{document}